# The road towards imaging a black hole – a personal perspective

Heino Falcke
*Radboud University Nijmegen*

**Abstract**
The images of the supermassive black holes Sgr A* and M87* by the Event Horizon Telescope (EHT) collaboration mark a special milestone in the history of the subject. For the first time we are able to see the shadow of black holes, testing basic predictions of the theory of general relativity. We are also now learning more about the fundamental astrophysical processes close to the event horizon that help to shape entire galaxies and even parts of our cosmos. The ultimate result was only possible due to a large collaborative effort of scientists and institutions around the world. The road towards these images was the result of a long sociological and scientific process. It started with early pathfinder experiments and a few simple ideas that were remarkably successful in predicting the basic observational signatures to look for. This was based on the premise that black holes are inherently simple objects. Here I describe this journey and some lessons learned from a personal perspective.

**Intro**
Imaging a black hole does not just happen. It is a result of a long scientific but also sociological process: pioneering work of a few and the collective effort of many, understanding the basic physics, long term visions, sudden revelations, technological game changers, continuously growing incremental insights, and in the end also some luck. Witnessing how such a process unfolds from the beginning was an interesting experience for me. Here, I try to tell the story from my personal perspective and mention a few of the lessons I learned along the way. This is not meant as an exhaustive and thorough scientific review of the topic, but uses a more narrative approach. An extensive and more popular description of the process and science can be found in our book "Light in the darkness" (Falcke and Römer 2021).

Looking back at my career it pays off to let yourself be guided by the big scientific picture first. As a student, I was drawn into physics by the fundamental questions about space, time, and matter that it addressed. Particle physics was at its peak and I was impressed by the large physics collaborations and the success of the standard model to rule our world view. However, it was also clear that particle physics was getting too big. The ability to make an impact was getting more difficult and the next big thing was perhaps decades away.

There was still one major unsolved question: What is the true nature of gravity – after all, it was the last force resisting assimilation into the standard model of physics. Was it a force at all? How would it relate to quantum physics? One of the most mysterious objects that encapsulated all these questions seemed to be black holes. Did they really exist in nature?

**Black holes: from mathematical curiosity to astrophysical paradigm**
Soon after Einstein developed his theory of general relativity (GR), black holes became the first solution of his field equations. In the trenches of World War I, Karl Schwarzschild (1916) calculated how spacetime would be curved around a point mass. Everybody understood that this was more or less a mathematical trick to make calculations simple and did not necessarily reflect a physical reality. Schwarzschild's solution was not called a "black hole" immediately[1] but it certainly was up for surprises. This space time metric was marked by a singularity at the center. Solutions to equations would explode impassibly into the infinite. Singularities in spacetime are points where world lines terminate, i.e. where light and information ceases to exist in an instant, and where space is infinitely stretched.

Equally interesting was a coordinate singularity that later became known as the event horizon: a virtual one-way membrane. Everything crossing this ultimate boundary can only travel towards the central singularity and never return – no matter, no light, and no information can leave this region. The event horizon poses a fundamental problem to modern physics: it separates a part of our universe that clearly exists, and where laws of physics should be at work, but which shields itself from measurement by any outside observer. Is not the ability to experimentally verify physical theory a defining feature of science? Is the inside of black holes still science, or is it beyond our science?

Einstein recognized these problems, but didn't really believe black holes could form in nature (Einstein 1939). They were merely a mathematical curiosity in his mind[2]. Einstein was soon shown to be wrong by Oppenheimer & Snyder (1939) and later by Penrose (1965). In fact, black holes can form naturally through the collapse of a massive star and they can grow to supermassive proportions in the center of galaxies by accretion of matter and mergers with other black holes. Of course, at the time it was not clear these objects would really exist, let alone that you realistically could see them.

The first hint that compact dense objects really could form, came from the discovery of white dwarfs and later of pulsars – ultracompact rotating neutron stars – by Jocelyn Bell-Burnell in 1967 (Hewish et al. 1968)[3]. Neutron stars form in the collapse and implosion of massive stars, but have a maximum mass limit beyond which they would turn into a black hole.

In the late 60ies and beginning 70ies X-ray astronomy led to the discovery of stellar mass black hole candidates in X-ray binaries (Webster and Murdin 1972). Radio astronomy turned up quasi-stellar radio sources, or quasars for short (Schmidt 1963; Hazard, Mackey, and Shimmins 1963). They were shown to be at a distance of billions of light years, radiating 30-100 times the luminosity of an entire galaxy from a region barely larger than our solar system, as one could derive from variability arguments.

How can one produce so much energy in such a small volume and how can one avoid that this powerful engine is blown away by the brute force of its radiation pressure? Supermassive black holes were the answer (Lynden-Bell 1969). Matter particles of mass *m* circling close to a black hole would approach almost half the light speed. This provides kinetic energies of a significant fraction of $mc^2$ which then could be turned into heat and radiation in dissipative processes. Accretion onto black holes can be more efficient than the process of nuclear fusion, powering the sun and the stars, which has an efficiency of ~0.7% $mc^2$.

---

[1] The complicated genesis of the name is traced in an article by Herdeiro and Lemos (2018).
[2] While I am not a historian of science, it does feel a little bit like the early discussions around the Copernican heliocentrism, where it was also debated whether it reflected only a mathematical or a physical reality.
[3] Anthony Hewish received the Nobel prize for this, Jocelyn Bell not. Fortunately, there is a still chance to correct this historical error.

At the same time, the enormous gravitational pull of black holes could contain the huge amount of energy in the very center of a galaxy against the huge outward facing radiation pressure – fulfilling the Eddington limit. Lynden-Bell and Pringle (1974) and Shakura and Sunyaev (1973) calculated the astrophysics of this process[4], where angular momentum forces the matter to rotate in a dense, intransparent accretion disk that stays geometrically thin and allows all locally produced energy to be radiated away by black body radiation (i.e., heat radiation).

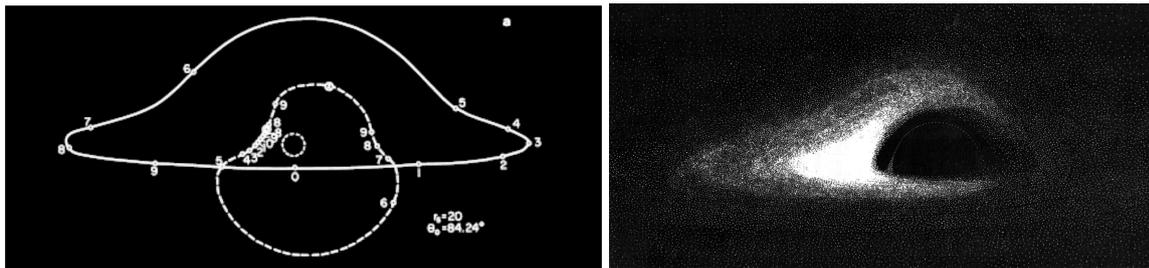

*Figure 1: Visible appearance of the orbit of a star around a black hole (Cunningham & Bardeen 1973, left) and appearance of a thin, intransparent accretion disk around a black hole (Luminet 1979, right)*

Soon theorists would calculate the appearance of such a black hole (Cunningham and Bardeen 1973; Luminet 1979, Figure 1). The black hole paradigm was set and our vision of black holes shaped. Today this is visible in popular science books and Hollywood movies like Interstellar[5]. While this picture is probably correct for the brightest sources in the universe, it is also dangerously deceptive and in fact misleading if one thinks about more normal black holes that the Event Horizon Telescope has imaged. Most of the black holes in the cosmos, do not radiate brightly like the few big "stars" that attracted all the attention in the 1970ies and their physics is different. It is exactly these feeble black holes in the universe that let us peer into their darkness and learn something about gravitational physics from them.

Radioastronomy soon found another curious phenomenon: black hole candidates are able to produce copious amounts of radio emission. That emission does not come from matter falling inwards, but from matter shooting outwards in the form of highly collimated plasma beams or "jets", flowing at velocities close to the speed of light and sometimes looking like the exhaust of a jet engine or a smoke trail from a chimney (Figure 2). The emission is due to synchrotron emission: high-energy electrons gyrating in strong magnetic fields and it this these magnetic fields that are strongly suspected to launch astrophysical jets.

Where there is smoke, there is fire, and these jets are also giant particle accelerators producing some of the highest energy particles in the universe. No wonder, since they are plowing with almost the speed of light through the interstellar medium of their host galaxies, producing all kinds of shocks and turbulence that is conducive to particle acceleration. The jets also produce X-ray and gamma-ray emission and also are likely sources of ultra-high energy cosmic rays (Blandford, Meier, and Readhead 2019) and neutrinos (IceCube Collaboration et al. 2018). The entire high-energy sky lights up with black holes and their associated jets! This provides an intriguing and somewhat unexpected link between particle physics and astrophysics, which has merged today into the field of astroparticle physics.

---

[4] The basic physics of such a thin disk was already calculated by von Weizsäcker (1948) for the case of forming stars and planets.
[5] A history of black hole imaging is presented in the PhD thesis by Emilie Skulberg "The event horizon as a vanishing point - a history of the first image of a black hole shadow from observation", Univ. Cambridge (2021).

Most of the energy of jets and accretion disks is generated very close to the black hole. This is the deepest gravitational well in the universe. The energy will either disappear into the black hole or be transported outwards and impact the stars and gas in the host galaxy via radiation from the accretion disk or the kinetic power of the jets. The more matter falls towards a supermassive black hole, the more energy is deposited inside the galaxy and in the medium in between the galaxies. Hence, what happens near black holes also effects the evolution of galaxies and even of the entire cosmos to some degree (Fabian 2012; Vogelsberger et al. 2020).

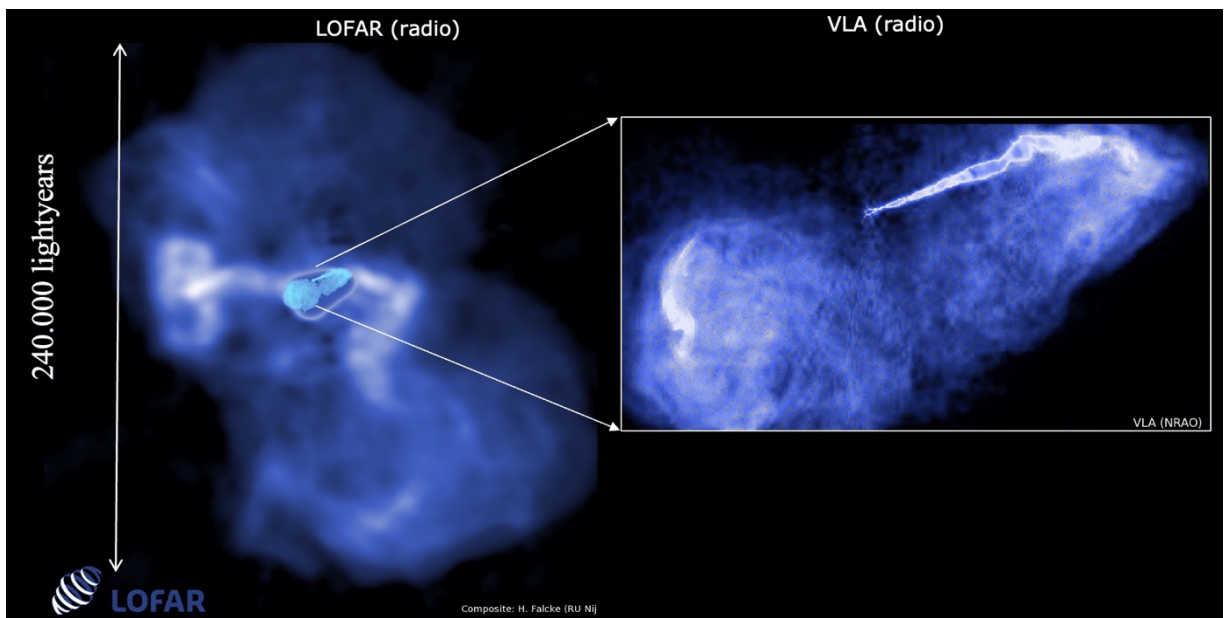

*Figure 2: The radio jet in the galaxy M87 (de Gasperin et al. 2012). Left is a radio image at 140 MHz made with the LOFAR radio telescope in the Netherlands and the insert on the right shows a higher resolution image made with the VLA in New Mexico (NRAO) at 15 GHz.*

Soon another thought kept astronomers busy: If there are supermassive black holes in distant quasars in the past, shouldn't there still be supermassive black holes in normal galaxies close to us? Lynden-Bell and Rees (1971) even speculated there should be one in the center of our own Galaxy, perhaps even marked by a compact radio source, like we see it in the hearts of many quasars. Sure enough high-resolution spectroscopy of gas and stars in the centers of nearby galaxies revealed the presence of dark compact objects there (Kormendy and Richstone 1995), including our own Galaxy (Genzel and Townes 1987). Radio astronomers (Balick and Brown 1974; Ekers et al. 1975) also found a compact flat spectrum radio source in the Galactic Center (Figure 3) not unlike radio cores seen in quasars, but much weaker. It was later called Sagittarius A* (Sgr A*)[6].

---

[6] Sagittarius A was simply the brightest radio source in the constellation Sagittarius. The star marks the "exciting" compact radio core in the Galactic Center (see Goss, Brown, and Lo 2003).

This object was henceforth target of many very long baseline interferometry (VLBI) experiments that tried to unravel its radio structure, going to higher and higher frequencies. VLBI combines radio telescopes around the world into a virtual earth-sized telescope, providing the highest imaging resolution in astronomy – the higher the frequency and the longer the separation of the telescopes, the smaller structures one can see. However, soon it became clear that the structure of Sgr A* was blurred by radio waves scattering off interstellar clouds (Lo et al. 1993; van Langevelde et al. 1992) - located somewhere between us and the Galactic center (Bower et al. 2014). This made it seemingly impossible to see the actual nature of the radio emission. Nonetheless, the hope never dies.

When I started with my PhD, I witnessed an ambiguous situation. Everybody was talking about black holes, but they were still seen as exotic and unproven: a ruling paradigm, but not a physical reality yet. Many sceptics demanded more proof.

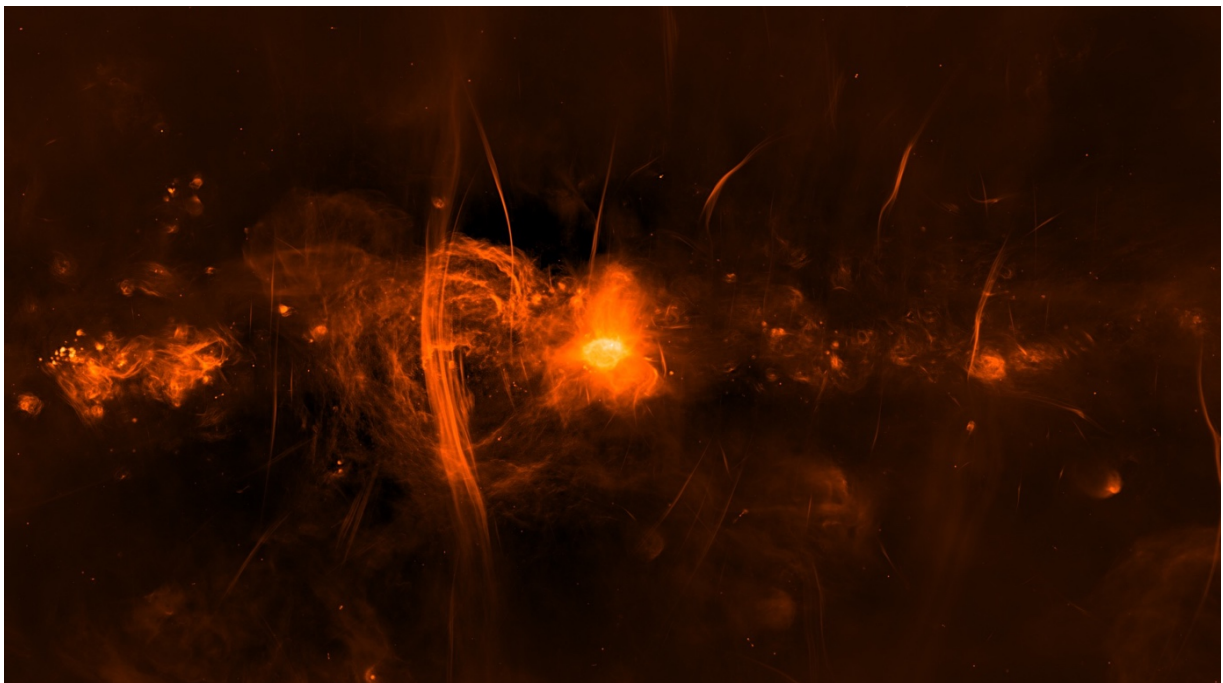

*Figure 3: Radio image of the Galactic Center region made with the MeerKat telescope array in South Africa at 1.3 GHz (Heywood et al. 2022). Sgr A\* is the point in the bright central splotch. The image has a width of 775 lightyears.*

**In the focus of new telescopes**
In the early 1990ies, several observational results involving Sgr A* caught our attention: new (sub)mm-wave detectors found surprisingly strong emission at the shortest wavelengths not yet resolvable by VLBI observations (Zylka, Mezger, and Lesch 1992). Was this dust or synchrotron emission from the black hole? A group at the Max-Planck-Institut für Radioastronomie (MPIfR) in Bonn soon produced the first ever VLBI observations of Sgr A* at a wavelength of 7mm, i.e. at the shortest wavelength and at the highest resolution ever achieved until then (Krichbaum et al. 1993). The image seemed to show some structure. Does one see a jet sticking out, calibration issues, or just blurring of the image by the interstellar medium again? Also, the Max-Planck team in Garching (Eckart et al. 1993) presented their first near-infrared speckle images of the Galactic center, revealing some faint emission. Was this the black hole or something else?

I had the benefit to work on theoretical astrophysics at the MPIfR in Bonn, i.e. in an institute, which specializes in observational radio astronomy. My supervisor, Peter Biermann, was an old school physicist in the best sense: able to calculate with logarithms, knowing the fundamental constants by heart, and able to quickly check the basic physics on a black board guided by observational reality – before one would go into an in-depth theoretical or numerical calculation. The sea of theoretical possibilities in astronomy is vast and data provides a shoreline to navigate.

However, what aspects of the data are believable? What technology can we reasonably expect in the future and what science can we expect from it. This you can only learn from the true experts in the field. Working with observers and instrument builders roots you in reality and opportunities. So, when these new observations of Sgr A* came about, it seemed clear to us: Something exciting is in the air and new observational opportunities abound - the hunt for the supermassive black hole in our Galaxy is on!

**Modeling Sgr A* - if you know one black hole, you know them all**
Surprisingly, there was no clear understanding of what the nature of the radio and submm-emission of Sgr A* is. Already a few months earlier we had investigated, what the accretion rate and luminosity of Sgr A* was, i.e. how much matter was falling towards the black hole and how much energy was dissipated into radiation. We concluded that the luminosity and the mass infall rate were extremely low (Falcke et al. 1993). We called it a "black hole on a starvation diet". Now we were wondering, whether the radio emission of Sgr A* could be related to the jets seen in much more powerful quasars and radio galaxies, but at much lower power levels.

At the time there was and still is a strange split in the community: either one would work on jets from black holes, mainly related to radio astronomy and later gamma-rays, or one would work on accretion disks, mainly working with optical or X-ray data. Communities in astronomy were traditionally determined by one's favorite observing wavelength – that would literally determine your view of the universe and of individual objects.

But aren't jets and disks not the expression of the same phenomenon and intimately linked? We later spoke a bit provocatively of the "jet-disk symbiosis" (Falcke and Biermann 1995). Our thought was: Black holes should be simple! From a grand perspective, their properties are only set by their mass, giving a length scale, and their accretion rate, setting the power scale. Spin and orientation of the black hole could modify the appearance further, but to a minor degree only. So, if powerful black holes have a jet, why not feeble ones like our own?

We postulated a fundamental coupling between the power of the inflow and the outflow: $P_{\rm jet} = q\dot{M}c^2$. Turn down the accretion rate, $\dot{M}$, by a factor of a Million, you get a jet power, $P_{\rm jet}$, that is lower by a factor Million (where the efficiency $q$ is of order of a few percent and $c$ is the speed of light). Calculating the radio flux using synchrotron emission and a simple conical source geometry (Blandford and Königl 1979; Falcke, Mannheim, and Biermann 1993), one can show that the radio flux scales non-linearly as $F_\nu \propto \dot{M}^{17/12}$ and that the radio size is roughly proportional to the wavelength.

This model very naturally explains the characteristic flat spectrum of compact radio emission not only in the Galactic Center, but of black holes in general. Indeed, we later hunted for this radio signature and found that it is ubiquitous in low-power galactic nuclei (Nagar, Falcke, and Wilson 2005).

When we applied this scaling to Sgr A*, we found that we only needed an accretion rate of $10^{-7} – 10^{-8}$ $M_{sun}$ (solar masses) per year to explain its faint radio emission and to reproduce its tiny size as found by the new VLBI measurements. This accretion rate is at least 8 orders of magnitude lower than what one derives for powerful quasars and about four orders of magnitude lower than the average accretion rate one needs to grow our Milky Way's central black hole over the age of the universe. 'Low', however, still means that it consumes the mass of the moon every few years.

Most excitingly, the model suggested that the higher the frequency, the smaller and closer to the black hole the radio emission should be. In fact, the (sub)mm-wave emission should come from a region the size of the event horizon. Indeed, if one looks at the Sgr A* spectrum, it cuts off at higher frequencies, i.e. shorter wavelengths. Was that a sign of the event horizon setting the smallest possible scale and suppressing higher frequencies? Would it perhaps be possible to see the black hole using that radio light?

A quick check of the size scales showed that the diameter of the event horizon for a Schwarzschild black hole, $2 R_s = 4 R_g = 4\,G\,M/c^2$, was just ~9 µas (microarcseconds). Astronomers measure sizes of celestials objects in angular scales. This can be turned into an actual size or width with basic trigonometry, only if one knows the distance of the object. One microarsceond corresponds roughly to an object with the width of a human hair in New York as seen from Nijmegen in the Netherlands.

The event horizon scale we calculated in the 90ies was for a mass M≈$2 \times 10^6$ $M_{sun}$ and a distance $D$=8.5 kpc (one kiloparsec, kpc, is 3262 lightyears = $3 \times 10^{21}$ cm). This was too small to be resolved by a VLBI array at 230 GHz, which would only reach about 25 µas resolution. The smallest scale resolvable by an interferometer is given by $\lambda/D$. Here, the wavelength $\lambda \sim 1$ mm was fixed by the astrophysics of the black hole and the maximum telescope separation $D$ was set by the size of the Earth – two parameters you cannot easily tune.

Moreover, black holes would likely be rotating and the event horizon could easily be a factor two smaller for a maximally spinning black hole. Another potential candidate, the supermassive black hole in M87 seemed even smaller at the time, by a factor three (Ford et al. 1994) - disappointing, but still intriguing and the thought kept nagging at me.

**A new vision**
A few years later[7], I realized that we had overlooked an obvious effect. Searching for another paper in a book, I accidentally stumbled over a rather unknown paper by Bardeen (1973), who calculated the appearance of a rotating black hole in front of a star. The 'black hole' he showed was five times larger than the event horizon. Only then it dawned on me that the strong gravitational lens effect would, of course, magnify the appearance of the event horizon in any image. Also the spin effects were much less than expected. If you illuminate a black hole, light has to pass on both sides of the spin axis. On one side it can "flow" with the rotation of space time and pass very close to the event horizon. On the other side, however, light will have to flow against the rotation of spacetime and be captured much further out (Figure 4). This means that spinning up a black hole will shift the location of the black hole centroid relative to the dark region, but it will not change in size much.

---

[7] It was probably around 1995, since the first mentioning of the possibility to image the black hole appeared in two conference proceedings: Falcke (1996) & (1997).

Unknown to me at the time, the basic equations for the light bending around a non-rotating black hole, had already been calculated by David Hilbert in 1916 (see Sauer 2009; von Laue 1921). He had shown that light passing tangentially at a distance of √27 $R_g$ would be asymptotically bent onto a circular photon orbit (at a separation of 3 $R_g$). Light rays coming any closer would end up in the event horizon, while light rays passing further away can still escape. Shining light at a black hole and looking at it would therefore reveal a sharp divide between darkness and light with a circular shape of radius √27 $R_g$. Including spin effects would only mean that the size could shrink by about ~8%. The effect therefore predicted a dark region of 25 µas in Sgr A*, just resolvable by VLBI at wavelengths shortwards of 1.3 mm. Also the expected blurring of the image by interstellar scattering was expected to become subdominant at this wavelength. Hence, it should actually be possible to start seeing "the black hole" with a realistic VLBI array at (sub)mm-waves!

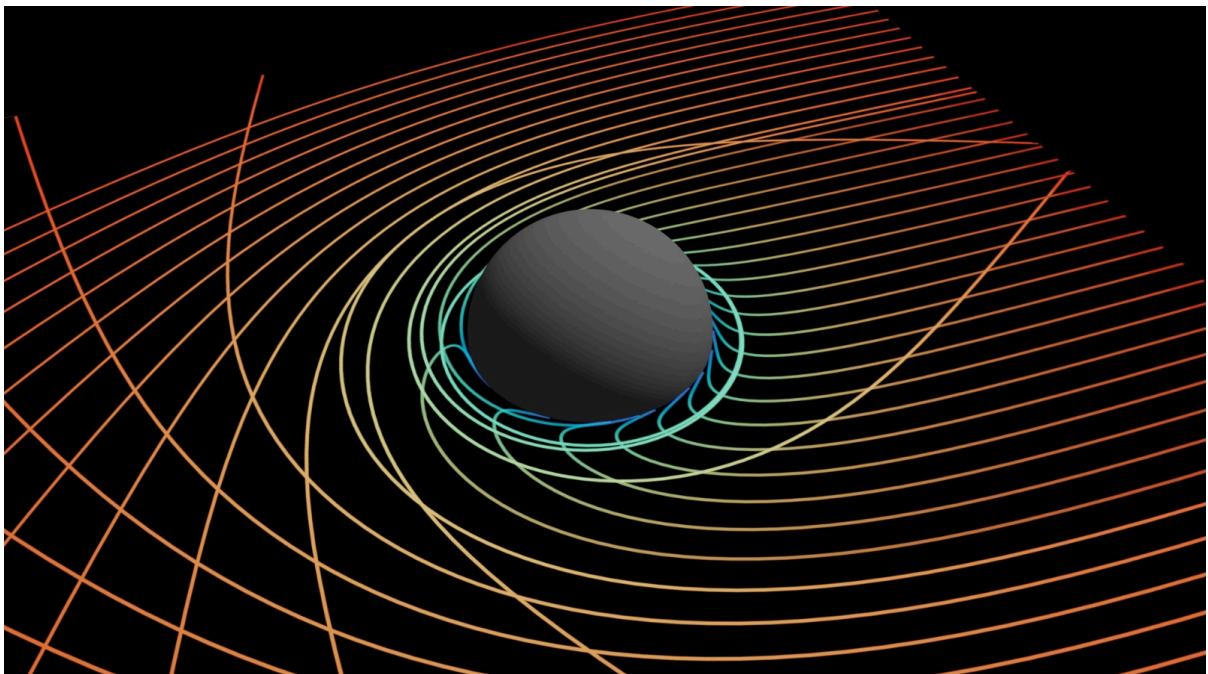

*Figure 4: Light rays in the equatorial plane of a rotating black hole. The colors indicate gravitational redshift (based on Bronzwaer and Falcke 2021a).*

That thought was a life changing revelation. Today most libraries are gone and so are many old books. Nowadays we pre-select our literature research with filters and bots or leave it to Google. I hope we still give luck a chance in science, by occasionally turning an unexpected page somewhere.

Going further from that point needed more than just a vision. Some hard work had still to be done to make a convincing case. Millimeter-wave telescopes were around already, but some were threatened to be closed, many were still to be equipped, and new ones had to be built. New opportunities appeared at the horizon with the planned ALMA telescope in Chile, the Large Millimeter Telescope (LMT) in Mexico, the upgrade of the Plateau de Bure Interferometer NOEMA, or later the Green Land Telescope (GLT). In the end, all would be used, but it took longer than expected.

**Casting a Shadow**

To make the case clearer we wanted to do a proper ray tracing of the situation in Sgr A* and make a theoretical image of the black hole. Images capture imaginations. We calculated the appearance of black holes for a wide range of generic astrophysical scenarios, where the black hole was embedded in an extended, transparent, glowing emission region as expected for Sgr A*. We explored various spins, as well as rotating, infalling, static, and outgoing flows and different radial emission profiles (Falcke, Melia, and Agol 2000b; 2000a, Figure 5 and Figure 6). While I had done general relativistic light radiative transfer for my Diploma thesis already, we here used a more modern ray tracing code that Eric Agol had developed to compute the radiation transfer of a thin accretion disk as a model for quasars[8].

In the calculations we broke with the standard picture of thin, intransparent discs. In the standard picture one sees in popular magazines, the dark region is dominated by a gravitationally lensed image of the hole in the accretion disk (Figure 1). Hence, in this case the key observable does depend significantly on the astrophysics. For example, seen under an angle, the hole would look egg shaped, since the inclined disc would obscure the black hole itself. It is hard to see a hole behind a wall.

However, in all our transparent and geometrically extended emission models we always saw the close-to-circular dark region caused by the light bending in the Kerr metric and absorption in the event horizon. We called that region the 'shadow' of the black hole[9], since one does not see the black hole itself, but rather the light it was blocking[10,11]. In contrast to the thin disk picture, the shadow also remains if the emission region is evacuated well beyond the shadow size (Narayan, Johnson, and Gammie 2019) – as long as the emission engulfs the black hole from all sides. In fact, the emission could even be at infinity. A simple picture to understand this is a wick in a burning candle, that leaves its dark imprint in the flame. Of course, for the black hole case one would not see any surface texture and the general relativistic black hole wick would appear larger by a factor 2.5 than it actually is due to gravitational lensing.

For the sample models that we ultimately published (Figure 5), we chose a red color scale, representing heat or molten iron. After all, we were showing radio emission that is invisible to the human eye, but comes from the extreme red part of the electromagnetic spectrum and is emitted by extremely hot gas. This was an artistic choice in the end, but it stuck.

In our simulations, the shadow was often surrounded by a bright ring of emission, later called a photon ring. Sometimes, only half or a quarter of the ring was clearly visible, due to relativistic beaming. Emission from plasma that is moving towards the observer will be amplified, plasma moving away will be de-amplified. This leads to a characteristic crescent shape for a rotating emission region (see also Kamruddin and Dexter 2013). Figure 6 shows a few models that we calculated at the time for different generic astrophysical situations.

---

[8] It was applied to the inversion of a microlensing light curve (Agol and Krolik 1999) as well as modeling accretion disks with non-zero torque at the inner edge due to magnetic fields (Agol and Krolik 2000). For the Sgr A* shadow application, it was extended to handle the 3D geometry.
[9] Incidentally, only a few weeks later a paper was published by (de Vries 2000), who introduced the same terminology completely independently.
[10] See Bronzwaer and Falcke (2021a) for a more extensive discussion of the nature of the shadow.
[11] Note, that the EHT now defines the shadow as the precise mathematical region encircled by the lensed photon orbit, which is different from the observed feature, which one could call observed shadow or central brightness depression.

The circumference of the shadow itself is mathematically precisely determined by the gravitationally lensed photon orbit, which reflects properties of the spacetime geometry. This makes it suitable for direct metric tests for any theory of curved spacetimes, like general relativity (see also Psaltis et al. 2015). Hence, we concluded that the shadow is a very robust test to probe key characteristics of black holes and their spacetime.

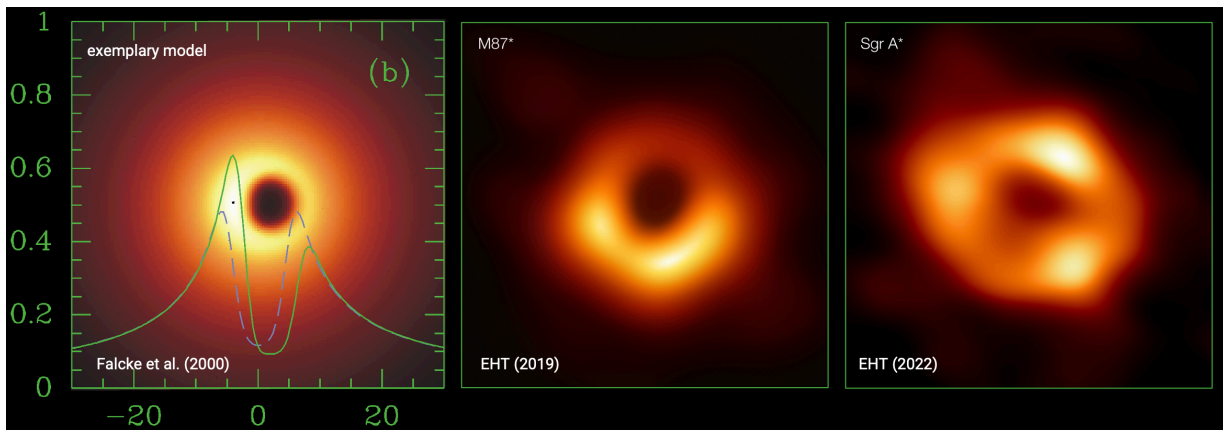

*Figure 5: An early model prediction of the black hole shadow in Sgr A\* from (Falcke et al. 2000a, left) compared to the images of M87\* and Sgr A\* published by the Event Horizon Telescope collaboration. This model was originally calculated at 0.6 mm wavelength for a rotating 2.6 Million solar mass black hole – as assumed at the time – with free-falling matter in an $r^{-2}$ emission profile. The size of the shadow is proportional to mass. Today's mass over distance (and shadow size) for Sgr A\* is a factor 2 larger. On the other hand, a telescope at 0.6 mm wavelength has a factor roughly two higher resolution than at 1.3 mm. Hence, in terms of resolution elements per shadow size the 0.6 mm model is nnow a better match to today's best parameters than the 1.3 mm model.*

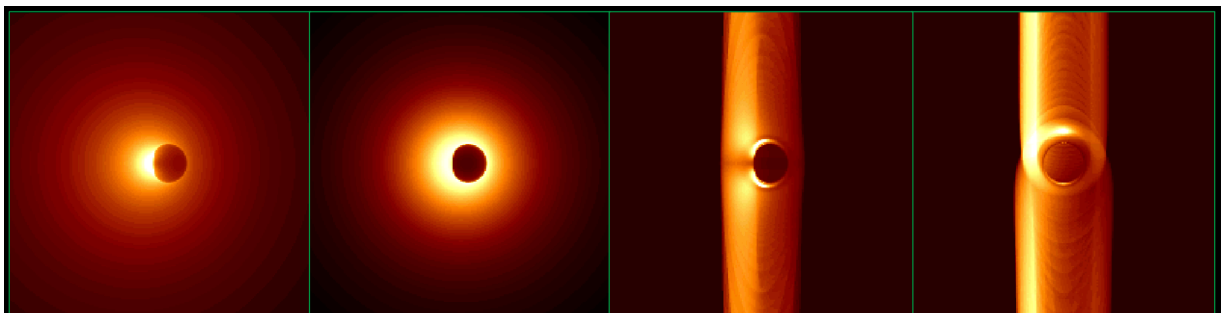

*Figure 6: Different shadow models showing a range of astrophysical situations. From left to right: a) Kerr black hole with spin a=0.998, inclination i=90°, $r^{-2}$ emission profile and "Keplerian" rotation; b) same as a) but for a free-falling non-rotating emission region; c) jet-like emission in a hollow cylinder with rotation and outflow for i=90°; d) same as c) but for i=45°. In all cases the imprint of the shadow is clearly seen. The asymmetry in the ring in a, c & d comes from Doppler beaming. (These images are from Falcke (2001) and in Falcke, Melia, and Agol (2000a), with the exception of panel c, which was only shown in the slides of my Ludwig-Biermann-prize lecture published in Falcke (2001)).*

Of course, the actually observed shadow will be slightly larger or smaller depending on the astrophysical model and the image resolution – there are never perfectly sharp gradients in nature and one will need some suitable calibration strategies to relate the observed shadow to the mathematical one. Today the EHT has developed methods to quantify the error associated with this and they are of order 10% - 20%, limited by the resolution of the experiment and uncertainties in the astrophysical model (Event Horizon Telescope Collaboration 2022a; 2022b).

**Probing the waters – testing the model**

Of course, probing the shadow in Sgr A* only would work if the emission is indeed what we thought it was. Hence, we first checked whether the submm-emission in the Galactic center was indeed synchrotron emission from the black hole or perhaps some unrelated dust further away, as some claimed. We organized what was perhaps the first multi-wavelength campaign with multiple telescopes looking at Sgr A*. This gave us a simultaneous broad-band radio spectrum and confirmed the non-thermal nature of the submm-bump. Again, a very simple analytic model of the emission showed that indeed the emission region should be horizon scale and becomes transparent at mm-waves (Falcke et al. 1998). Already in this paper we strongly advocated imaging the event horizon at (sub)mm-waves.

Soon thereafter, the first prototype VLBI observation of Sgr A* at 1.3 mm was published by a small group at the MPIfR group in Bonn (Krichbaum et al. 1998). It used the two IRAM telescopes on Pico Veleta and the Plateau de Bure. Two telescopes are not enough to make an image, but at least a size could be measured. The inferred value was consistent with today's measurements, however, the error bar was large and the conclusions unclear. Also experiments at 3 mm got better, showing that the emission was compact indeed (Doeleman et al. 2001), but it still was not clear whether interstellar scattering was fooling us.

A few years later, we were able to get much better data at longer wavelengths and finally succeeded to derive an intrinsic size of Sgr A* after subtracting the interstellar blurring at various wavelengths (Bower et al. 2004). Indeed, the actual source size was decreasing with wavelength, consistent with our earlier jet models. Extrapolating, one could show that event horizon scales would be reached at around 1mm wavelengths (Figure 7).

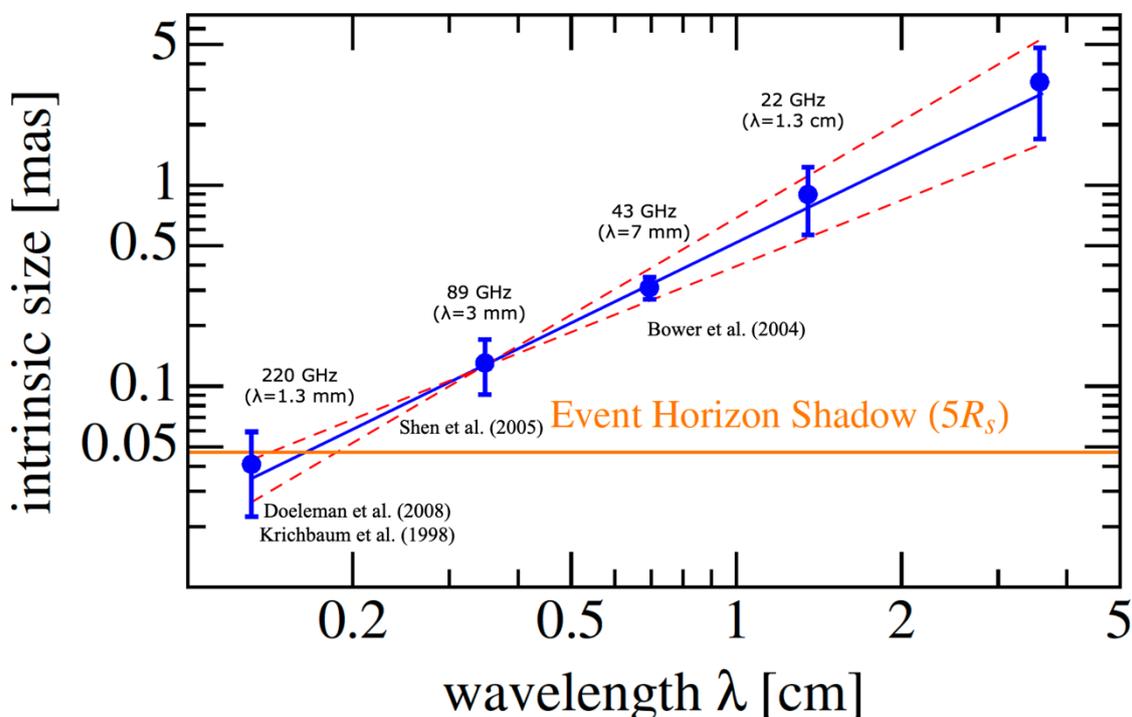

*Figure 7: Intrinsic size of Sgr A\* - after subtracting an empirical scattering law broadening the image - as function of wavelength, prior to the EHT measurements. At 1.3 mm wavelength the emission approached the size of the shadow. (Adapted from Falcke and Markoff 2013). A more recent size determination at 3mm can be found in (Issaoun et al. 2019; 2021).*

**Building momentum**

In 2004 we organized a special session at a conference in Green Bank dedicated to the 30th anniversary of the discovery of Sgr A*, where Geoff Bower, Shep Doeleman and myself spoke[12]. The discussion was geared towards preparing future experiments and at the end of the session, there was strong support in the community for conducting the experiment. Haystack observatory started preparing a program to develop broadband digital equipment necessary to increase the sensitivity of the experiments and to improve the ability to correct atmospheric phase fluctuations, that was hampering proper imaging. Also, the newly developing ALMA telescope project seemed to embrace black hole imaging as a potential science case (Shaver 2003). Unfortunately, this mode was not implemented from the start and had to be added later (Matthews et al. 2018).

In a series of telecons the three of us discussed a potential road towards the next generation of experiments. In a meeting on March 3, 2006, I summarized in the minutes the state of the discussion:

"Either we have a loose collaboration or a more formal one? In a closer collaboration one would sign an MoU[13] defining the structure (collaboration board, speaker and co-speakers, publication policy, financial matters, collaboration meetings, working groups on experimental issues and simulations, data reduction and interpretation).

The goal of the consortium will be to not only image the shadow of the black hole once but to continue the experiment in the future. We want to turn Sgr A* into a precision laboratory to directly measure and image all kinds of general relativistic effects that cannot be seen anywhere else in the universe."

I was advocating a more formal collaboration, since I had started to work in astroparticle physics and had seen how larger physics collaboration worked from the inside. However, it would take a little longer for it to take shape. Europe was losing its leadership at the time. Germany was leaving the Heinrich-Hertz-Telescope in Arizona, which was later renamed to submillimeter telescope (SMT), and the Netherlands the James Clerk Maxwell Telescope (JCMT) in Hawaii. The main argument was the new Alma and APEX-Telescopes coming up, which needed funding and attention. Astronomers had the reputation of always building new telescopes without giving up old ones. Just at the time when (sub)mm-wave astronomy was entering a new phase, with ALMA as a major new project, we faced the danger of losing the smaller telescopes, necessary to do a VLBI experiment!

Doeleman et al. (2008) organized a series of experiments with telescopes available in the US, which eventually succeeded with the first measurement at 1.3 mm with three telescope sites, including those two telescopes, plus the Combined Array for Research in Millimetre-wave Astronomy (CARMA) in California. The result was more robust and had smaller uncertainties then the experiment from the 90ies. Moreover, the mass measurements of Sgr A* had become more accurate and precise due to the measurements of stellar orbits around the object by the groups of R. Genzel and A. Ghez (Schödel et al. 2002; Ghez et al. 2005). The inferred mass had doubled and the inferred distance had shrunk, compared to the 90ies.

---

[12] See http://www.aoc.nrao.edu/~gcnews/gcnews/Vol.18/editorial.shtml & http://www.aoc.nrao.edu/~gcnews/GCconfs/SgrAstar30/. An early discussion of VLBI was organized for the Tucson 1998 meeting, which has been documented in Zensus & Falcke (1999).
[13] MoU = Memorandum of Understanding, a milder form of a contract

This meant that the expected shadow had increased significantly and should be about 50 μas in diameter now. The new VLBI experiment then claimed $36^{+16}_{-10}$ μas. Clearly, there was event horizon scale structure and these experiments paved the way for the future EHT.

In Europe we lobbied to get the idea of imaging black holes with submm-VLBI into the Astronet strategic plan for astronomy in Europe[14]. This did not deliver any money immediately, but came in handy when applying for major funding later. Shep Doeleman organized a meeting at the AAS-Meeting 2009 in Long Beach in preparation of the US decadal review (Doeleman et al. 2009). I organized a meeting at ESO targeting the European community and followed it up with a White Paper (Tilanus et al. 2014; Falcke et al. 2012).

In one of these famous conference coffee breaks at the AAS meeting in 2009, I talked with Shep Doeleman and suggested to choose a more interesting name for the experiment and we came up with "Event Horizon Telescope". A name was born[15], but there was no consensus yet, what it meant. Many meetings and telecons still had to follow.

Of course, there were also tensions and one could write an extra book about those. Not all of them were always constructive, but in the long run tensions also hold you sharp and can move things forward. In 2013 I succeeded together with two colleagues (Rezzolla & Kramer) in getting the 14 M€ Synergy Grant "BlackHoleCam" of the European Research Council (ERC), which enabled us to fund not only hardware, but also data analysis, theory, and even pulsar research (Goddi et al. 2017). This was the largest single grant ever given towards the EHT, that also integrated all these components into one large project. Fortunately, the ERC rules gave us a lot of freedom to spend the money flexibly in coordination with the other groups around the world. I was lucky to find a very able project manager, Remo Tilanus, who later also became the EHT project manager and had a lot of freedom to use the funds for wherever it was needed to quickly prepare for the experiments. This is the strength of an ERC grant. It also helped to attract and build a stronger network in Europe. A couple of months later also a major grant by the NSF was granted in the US with Doeleman as PI and many US colleagues contributing and in 2015 the East Asian observatory rescued the JCMT on Hawaii.

In 2014 the preliminary structure of the EHT was decided in a meeting at the Perimeter Institute in Canada and included 13 stakeholder institutes from Asia, Europe, and the Americas, all bringing in significant institutional resources and forming an initial board. The EHT indeed became a comprehensive collaboration with the ambition to cover all aspects: providing hardware, operating an array, calibrating and imaging data, analysis, and interpretation. This would later be reflected in the fact that the EHT eventually published more than just a single image, namely a comprehensive series of extensive papers covering all these aspects.

The final Memorandum of Understanding (MoU) was only signed in 2017 after additional 50 interim board meetings, actually well after we had conducted our first full-scale groundbreaking experiment! However, we operated under the emerging MoU already from 2016 onwards. Shep Doeleman became project director, I chaired the science council. Working groups were installed and a managing team was put together. The EHT leadership was from different regions of the world, but not as diverse as one might want. The consortium was young and had to find its way. Procedures and policies had still to evolve. The entire process reminded me of ancient nation building, where different tribes come together to form nations with an emerging legal and political system – fortunately, it progressed less bloody than in the old days, but not without stress and anxiety.

---

[14] https://www.eso.org/public/products/books/book_0028/
[15] Interestingly Makoto Miyoshi & Kameno 2002 and Miyoshi, Kameno, & Falcke 2003 also talked of an array called Event Horizon Telescope, but those conference proceedings were no longer on our minds five years later and I had completely forgotten them until I was interviewed by a historian of science from the Max-Planck-Institut für Wissenschaftsgeschichte (MPIWG) in Berlin.

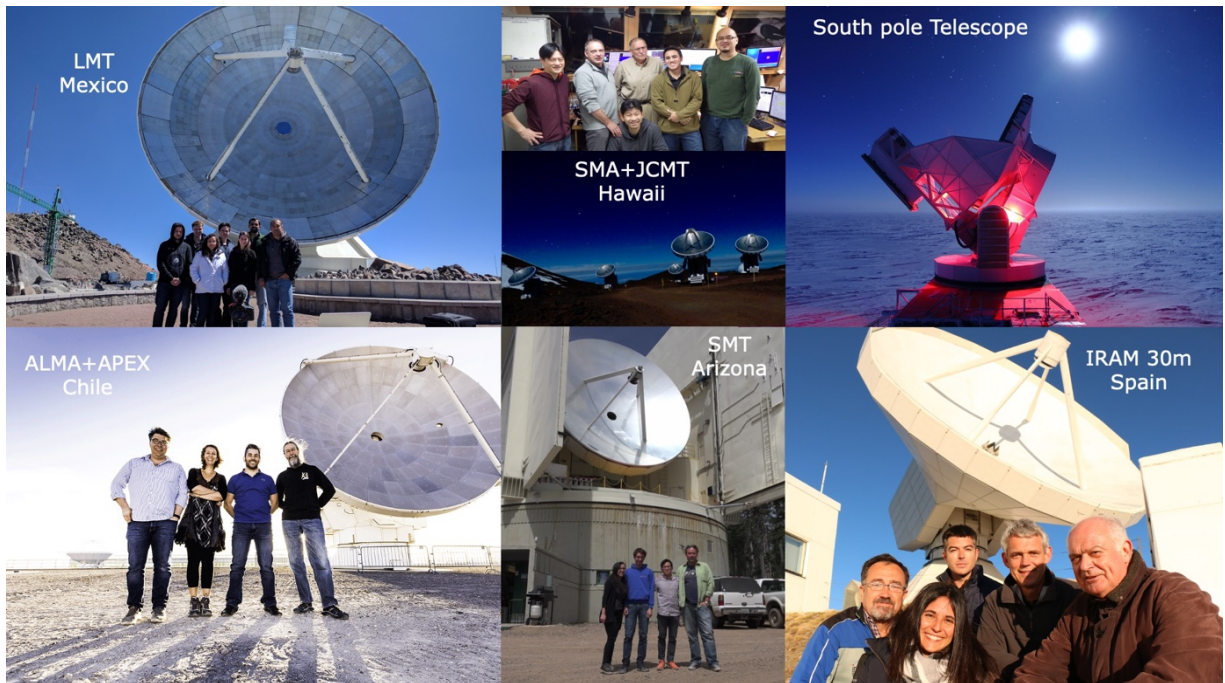

*Figure 8: Some of the telescopes and parts of their crews that were involved in the 2017 observing campaign of the EHT.*

**Feeling lucky**
In foresight and in hindsight the entire process seemed straightforward and unavoidable to me, but in between it was hard fought and debated throughout. Luckily enthusiasm carried us through and there was also the necessary luck.

In 2017 we carried out our first large scale attempt at an experiment with 8 telescopes on 6 different high sites (Figure 8). The crews were mixed from different institutes. Our group had sent personnel to all mountains, except South Pole[16], and met there scientists and observers from other stakeholder groups in the EHT.

I still remember looking at the marvelous blue sky at the IRAM 30m telescope at Pico Veleta, all the stories of weathered-out observations in the back of my mind. Millimeter waves are absorbed by water vapor in the atmosphere and you do not want to have a cloudy sky in between you and the universe. Amazingly, the weather was excellent around the globe during most of the ten day observing run and also all the equipment worked.

Fortunately, also the universe cooperated. A few years earlier one group (Gebhardt and Thomas 2009) claimed that the black hole in M87 was actually 3 times larger than thought in the 1990ies. That too would make it resolvable by the EHT, but another group disagreed (Walsh et al. 2013). We did not really know for sure what was right until we saw our data in May 2018. This was not an image, but first calibrated visibilities. I'll never forget the moment when my PhD student Sara Issaoun showed it to me "hot off the press". It was shouting at us: There are two black holes, where you can see a shadow!

---

[16] Which was equipped in a heroic effort by the University of Arizona and staffed by "winterovers".

That changed our plans. Rather than grappling with the variable, scatter-broadened image of Sgr A*, we were able to develop and exercise all our procedures on a more stable source that was repeatedly and consistently observed over four days. M87* is intrinsically 1500 times larger than Sgr A* and hence it takes 1500 times longer for gas to rotate around it – 12 days rather than 12 minutes, even at the speed of light. 25 years of planning suddenly changed within a few weeks.

Repeatability and duplication became a guiding principle in the analysis. Every step in the analysis was done by at least two independent groups, using independent software. This principle too took a while to emerge. A useful warning sign was the publication of the BICEP2 results a few years earlier. The experiment published highly celebrated ground breaking results, that had to be withdrawn within a year because of a single insufficient step in the data analysis.

However, an equally important factor was the natural competition in the end. Even in a collaboration that publishes with an alphabetically sorted author list, everybody wants to show that their algorithm is as good or even better than the other. It is also a questions of independence. Within our BlackHoleCam project we had the ambition to be able to conduct all aspects of the data analysis independently ourselves or in collaboration with other European groups. Similar dynamics evolved in other regions represented within the EHT.

In a way, this competitive collaboration principle was even reflected how the first results were presented: not in one central press conference, but in parallel ones in Brussels, Washington, Tokyo, Taipei, Santiago de Chile, and in additional satellite events. As a consequence the release of the first set of papers and images became a truly global event. For me Brussels was a natural choice to recognize the contributions of the European taxpayer. It was an enormous privilege that I could present the image to the public after all these years, but it was also a liberation after the intense months of collaborative work in secrecy. To me this was the moment when the image became reality. It didn't belong just to us anymore, but became global property. We had taken the image with the world, we gave it back to the world, and the world embraced it like none of us ever expected. More than four billion people were exposed to the image in the end (Christensen et al. 2019). Of course, a large fraction of those people probably forgot it also within a few weeks. Still, the impact on the public was huge and scientists around the world watched the various press conferences live.

The competitive collaboration principle worked very well in the end, if one looks at the final results. No other interferometric data set has ever been vetted so extensively by such a large and dedicated group of experts. It also has been reproduced by a number of groups in the meantime (Carilli and Thyagarajan 2022; Arras et al. 2022; Lockhart and Gralla 2022; Patel et al. 2022)[17] as well as by the EHT itself (Event Horizon Telescope Collaboration et al. 2021). However, competitive collaboration can also be exhausting and stressful. This requires proper management and policies, but those were just emerging during the process.

When the polarization data of M87*, revealing the event horizon scale magnetic field structure (Event Horizon Telescope Collaboration 2021), and finally also the image of Sgr A* (Event Horizon Telescope Collaboration 2022) were released, we already had a better sense of how the process would work and in the running up we tried to lower the pressure by taking more time before we published. It still was stressful. However, the result is something we can all be proud of. Behind the press release and the image, was an entire book of scientific literature that defined an entirely new level of black hole science, opening many new avenues of research.

**Physics or Astrophysics?**
One question still asked sometimes is: how much is this image really a test of general relativity, i.e. a measure of "true physics" rather than a result of arbitrary astrophysics. How reliable are the conclusions?

First of all, astrophysics has become a key pillar of today's physics. If you want to derive extreme physics from the extremes of the cosmos, you need to understand extreme astrophysics. Understanding the astrophysics surrounding black holes is equally fundamental as understanding the inside of the atomic nucleus or the nature of quantum interactions.

Secondly, the shadow is a fundamental property of a black hole in the Kerr metric and it does not depend much on the astrophysical model, if you are in the regime of transparent extended emission. Its size is accurately predicted by GR and scales linearly with mass. This is a very unusual property for any macroscopic object. In objects made out of normal matter, the mass would scale with the volume, i.e. with the $3^{rd}$ power of the radius and not linearly. This difference in scaling we can now test with the EHT alone over three orders of magnitude (Event Horizon Telescope Collaboration 2022b).

However, we can go even further than that. Given that gravitational waves are also subject to gravitational optics one can show that the gravitational wave signals measured by Ligo/Virgo of stellar mass black holes are emitted in the same region as the EHT radio emission and is probing the same components of the spacetime metric with roughly the same error bar of ∼20% (Psaltis et al. 2021). This allows one to confirm the scale invariance of general relativity over eight orders of magnitude very accurately, where the error bars become almost irrelevant.

---

[17] One paper claims to not reproduce the ring, but the EHT collaboration found the analysis to be fundamentally flawed (see https://eventhorizontelescope.org/blog/imaging-reanalyses-eht-data).

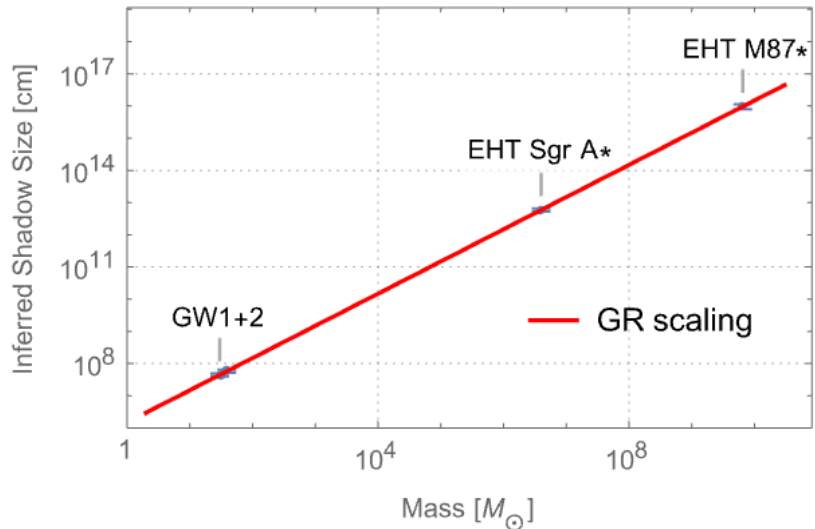

*Figure 9: Inferred shadow size of the supermassive black holes Sgr A\* and M87\* as measured by the Event Horizon Telescope and the shadow sizes inferred from two gravitational wave sources (here labelled GW1+2) measured by Ligo/Virgo as derived in Psaltis et al. (2021). This is a variant of a figure presented in Event Horizon Telescope Collaboration (2022b).*

Thirdly, astrophysical theory isn't arbitrary. We have exquisite models these days. Already for some time now numerical simulations have become a tool of their own, bridging pure theory and experimental physics. From the design of airplane engines to the prediction of climate change and the simulation of black holes, numerical simulations help us to study, predict, and observe nature. Astrophysical jet formation, for example, was long considered a mystery. Today jets form naturally in our simulations (McKinney and Gammie 2004).

Here, the EHT has brought the astrophysical study of black holes to a new level. For the first time in this field, theorists have cooperated on a large scale, checked their codes against each other (Gold et al. 2019; Event Horizon Telescope Collaboration et al. 2019b), agreed on common procedures, explored a wide range of parameters together in massive simulation efforts, and compared it to the actual data. That, together with a large multi-wavelength campaign (EHT MWL Science Working Group et al. 2021), narrowed down the parameter range of possible models significantly. Simulation software and supercomputer time have now become an equally valuable resource as telescopes or data analysis software.

Like in particle physics, the EHT also has developed a complete end-to-end simulation pipeline that starts at detailed black hole simulations (Event Horizon Telescope Collaboration et al. 2019b) to predict the physics, followed by advanced detector simulations (Roelofs et al. 2020), which capture most uncertainties in the measurement process, and dedicated parameter extraction methods (Event Horizon Telescope Collaboration et al. 2019c).

**Black holes are simple**

One other important realization is that we are in a regime of theory that can actually be tackled in a reasonable way. In fact, simulating a black hole is in principle easier than simulating our sun. Of course, we still have much less detailed information about black holes than about the sun, hence, there is less data to explain, but in principle this should be true, since black holes are described by much fewer parameters.

For example, our early jet model calculations could be done on a piece of paper using some very basic arguments, like energy and mass conservation, assuming an isothermal, supersonic, magnetically dominated turbulent outflow, which was launched near the black hole. Surprisingly, If we take today the most advanced supercomputer models of magnetized plasma flows around black holes, they show very similar scaling behaviors (Mościbrodzka and Falcke 2013; Mościbrodzka, Falcke, and Noble 2016). The best fitting model found by a wide exploration of these models by the Event Horizon Telescope collaboration (Event Horizon Telescope Collaboration 2022c), matches pretty much the pen-on-paper calculations I did for my PhD two decades earlier. It is surprising that we can explain such a wide variety of properties with models that start from first principles.

Another important aspect of our understanding of these starving supermassive black holes came from the study of their accretion flows (Narayan, Yi, and Mahadevan 1995). The key point here is that at low inflow rates, accretion disks cannot radiate all the locally dissipated energy from the inflow away. Protons are more massive than electrons, so carry most of the kinetic and thermal energy of the flow, while the lighter electrons, can wiggle faster and produce radiation more easily. The plasma is so tenuous, that protons and electrons hardly interact via electrostatic (Coulomb) forces, hence there is a disconnect between the two. Electrons cool more rapidly via radiation and energy transfer from the bulk of the plasma stored in the protons to electrons is inefficient. Therefore electrons could have a lower temperature keeping the heat in the protons. Rather than radiated away by the electrons, the energy is now advected towards the black hole by the protons in an advection dominated accretion flow (ADAF).

The temperature of the plasma near the black hole is then roughly given by the virial temperature T ~ ½ $G M m_p/k_B R_{ISCO}$ ~ $10^{12}$ K for $R_{ISCO}$ ~ 6 $G M/c^2$, $m_p$ is the proton mass, $k_B$ the Boltzman constant, and $G$ the gravitational constant. This is constant for black holes of all masses, as the escape or virial speeds are all of order $c$ in black holes! The original ADAF model tried to explain the entire Sgr A* emission without a jet component, but underproduced low frequency radio emission and hence demanded much higher accretion rates. Explaining the entire radio spectrum with only the inflow requires extra assumptions about a radial dependent acceleration on non-thermal particles (Özel, Psaltis, and Narayan 2000).

A jet, on the other hand, naturally reproduced the flat-spectrum radio emission. Here one needs to assume that as the plasma enter the jet, electron and proton temperatures come into balance again (Yuan, Markoff, and Falcke 2002; Mościbrodzka, Falcke, and Shiokawa 2016) – potentially via magnetic fields, which should be dominant in jets. Hence, the electrons can be hotter in the jet outflow, radiate more efficiently than in the inflow, and dominate the overall radio emission, even though the jet is less dense than the accretion flow. This also naturally explains why the radio cores observed by VLBI experiments reach +brightness temperatures of $10^{10}$-$10^{12}$ K, which is close to the virial temperature near the horizon.

The microphysics of the heating, however, still remains the biggest uncertainty from a physics point of view: these processes happen at small scales not captured by the simulations and hence need to be treated phenomenologically. For this reason the EHT explored a wide range of models from disk- to jet-dominated emission.

At these high temperatures plasma also becomes completely ionized, which means that all the elements are largely dissolved into just protons and electrons, giving the plasma itself a rather simple and universal composition.

Finally, there are magnetic fields in the flow which will be amplified via compression or other processes and can launch the jet outflow. This field can, however, only reach a maximum level set by the accretion rate. Beyond that level accretion flows become magnetically dominant and disrupt the accretion process (Tchekhovskoy, Narayan, and McKinney 2011). That maximum is set by the accretion rate and the size of the black hole: $\eta \dot{M} c^2 \sim B^2/4\pi \times 4\pi R_{ISCO}^2$, where the first term sets the total energy scale times an efficiency factor. This determins a rough scale for the magnetic field $B \propto \dot{M}^{1/2}/M_{bh}$ (Falcke and Biermann 1995).

The consequence of all of this is that despite the perception that black holes are very exotic and complex, they are to first order rather simple objects if they accrete matter at low rates. The black hole itself has little personality, just mass and spin, and neither has the plasma, which has a simple composition always moving with the same speed and the same temperature near the inner edge of black holes. Also the emission process is rather fundamental: largely relativistic electrons gyrating in the magnetic field. Most properties scale to first order with mass and/or accretion rate.

**Connecting the scales**

Putting this all together, one can arrive at some simple scaling laws for the non-thermal emission spectrum of these underfed black holes. This has led to the "fundamental plane of black holes" (Merloni, Heinz, and di Matteo 2003; Falcke, Körding, and Markoff 2004) which connects the optically thick (typically radio) and optically thin part of the emission spectrum (somewhere from submm-waves to X-rays) for black holes of very different masses and accretion rates, holding over at least 8 orders of magnitude. The turnover over point in the spectrum is roughly the frequency, where the emission is generated near the black hole and it scales as $\nu_t \propto \dot{M}^{2/3} M_{bh}^{-1}$ (Figure 10).

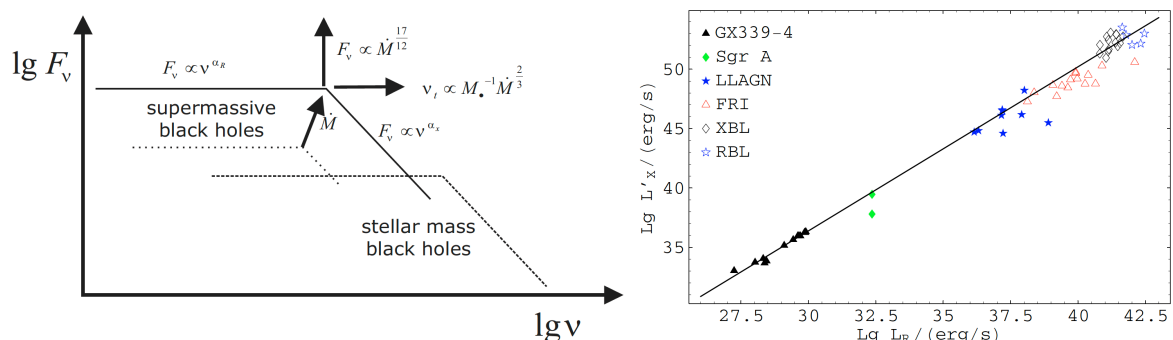

Figure 10: Simplified radio-to-X-ray spectrum of the synchrotron component of the radio jet emission in supermassive (left, solid line) and of stellar mass black holes (left, dashed line). The radio flux, $F_\nu$, is proportional to the frequency $\nu$ with a flat power law ($\alpha_r \sim 0$) in the intransparent part and a steeper power law $\alpha_x \sim -0.7$ in the transparent part. The turnover frequency where the spectrum changes and which marks emission close to the black hole, scales as $\nu_t \propto \dot{M}^{2/3} M_{bh}^{-1}$. On the right one sees radio and x-ray emission of low-power black holes, where the x-ray flux is corrected for the different mass of the objects following the basic scaling laws on the left (Falcke, Körding, and Markoff 2004). This connects black holes of very different masses and accretion rates. For Sgr A* the flare and quiescent x-ray flux is shown. Without the mass correction stellar and supermassive black hole would not fall on the same line.

Hence, which frequency one needs to choose to observe a black hole, also depends on its mass and accretion rate. The two black holes that the Event Horizon Telescope has imaged, M87* and Sgr A*, differ by a factor 1500 in mass and ~100.000 in accretion rate. However, since $100{,}000^{2/3} \sim 2000$, the scaling conspires such that both can be observed more or less at the same frequency. This is basic physics as much as it is total coincidence and luck again.

Also the fact that M87* is 1500 time larger, but also 2000 time further away than Sgr A*, giving both objects roughly the same angular size on the sky, is a complete coincidence. This makes it possible to observe both with the same experiment at the same frequency and verify the basic findings with two independent and very different objects.

A counterexample is the supermassive black hole in the radio galaxy Centaurus A, which was also observed by the EHT (Janssen et al. 2021). Its black hole is too small to be resolved and one would need THz frequencies to see it. Indeed, rather than a ring, the EHT saw a pair of scissors, i.e. the edges of the jet further out – which is interesting in its own right.

To see the black hole in Cen A one would need to go into space. Stellar mass black holes on the other hand, would need an X-ray interferometer in space (Cash 2005), since they are even smaller and the characteristic emission frequency is much higher.

Still, it is important to realize, that while the EHT has only imaged two black holes, our models already cover a wide range in astrophysical parameters. Given that the physics scales well – at least to first order, we can use the knowledge we gather from these two sources to understand a rather wide range of black hole systems in the universe – we do not need to image every black hole in the universe to understand them all! In fact, when we developed our models of Sgr A* in the early 1990ies, we always had an eye on the fact that whatever we did should not be particular to one sources, but should have the potential to apply to all. That guided us into the proper direction in the end.

The scaling laws include stellar mass black holes, whose properties can be compared well to their supermassive counterparts. They also allow us to also study long-term temporal evolutions in a human life time, that is inaccessible in individual supermassive black holes as it would take millions of years there. These stellar mass black holes also show the limitations of this approach: once one goes to the highest accretion rates and the brightest sources, the accretion flow becomes much more complex and unstable, they may become intransparent, and jet formation can be quenched (Fender, Belloni, and Gallo 2004). This more complex behavior at the bright end of black hole luminosity function is potentially caused by radiation pressure and opacity effects, which are much more difficult, but not impossible, to compute in the long run.

So, the irony of this story is that the first black holes that were discovered and which were the target of many theoretical and observational investigations over the past decades are also the most difficult to understand, while the unspectacular fainter ones may hold the true key to understanding black hole physics.

**Some conclusions**
We have witnessed an extraordinary period over the past 3 years. For the first time, we have seen images of black holes. There are only two objects, but they are representatives of very different types of galaxies and astrophysical conditions. We can now safely assume that the centers of galaxies do host supermassive black holes – from the first quasars to ordinary run-of-the-mill galaxies like our own.

The black hole images generally match earlier predictions made within the framework of black holes in general relativity remarkably well. This confirms that the shadow is indeed a characteristic and robust signature of these objects.

For the foreseeable future Sgr A* and M87* will also be the last black holes we see – all the others are too small or too far away. However, in contrast to gravitational wave emitting black hole mergers, these black holes will stay around. We can observe them with ever better techniques and at all possible wavelengths. With the EHT we can also study their temporal behaviors, maybe make a movie of their dynamics, and ever refine our models that can be tested against observational data. This is what you expect from a good physics experiment!

In the future we can turn towards space interferometry missions to image many more black holes and at much higher resolution and image quality (Roelofs et al. 2019; Palumbo et al. 2019). This can bring at least another order of magnitude or more improvements in the measured shadow size and rule out a much larger range of parameter space of alternative models. We can also understand black hole astrophysics in detail, e.g., fully understand jet launching, measure black hole spin much more precisely, and verify such effects as spin energy extraction from black holes through magnetic fields, via the Blandford-Znajek (1977) effect.

Looking back at the past 25 years it was a great privilege to witness how all of this unfolded. Getting to this point took an enormous effort by many scientists. Not one person alone, not one institute alone, not one country alone could have realized this vision in the end. Seeing how this collaboration emerged was almost as interesting as seeing how the science evolved.

It was a joy to see the scientific process unfold. Sometimes physics can be amazingly simple and can make decent predictions. It is always an almost magic experience, when predictions actually come true.

However, science is also a sociological process, that could have gone wrong at many turns. Here too it helps to keep the bigger picture in mind and make sure that big science problems are addressed in a big way. Insisting on scientific rigor is an important principle. It also helps to build competitive aspects into a collaboration. This competition needs rules, policies and mediators to keep the pressure at a manageable level. You cannot develop this overnight.

Also a proper reward system needs to be in place, that recognizes those who do pioneering work, as well as those who bring the work to fruition. In fact, seeing your own PhD student presenting a major science results at a press conference, can feel as rewarding as presenting it yourself. Having a sophisticated system of authorship rules that distinguishes alphabetic ordering of big collaboration papers from papers that are led by a small group for the collaboration or from method papers that are published by a smaller set of scientists is a useful tool. Not everything needs to be alphabetic, individual contributions should be recognized.

Finally, science needs money. A good idea is not enough. On the road towards discovery you need real estate and resources: telescopes and instruments, supercomputers, software and people. From my European perspective the ERC is really exceptional here. This is a very competitive funding program setup by scientists for scientists, which is strictly bottom-up, without any political interference or steering. Only excellence counts and bureaucracy is kept at a minimum (though it has been growing in recent years). It is the scientists who are entrusted with the success of their projects. They are relatively free to allocate their resources where they are actually needed. This strategy really pays off.

Of course, we also profit from the diverse and relatively stable set of institutional funding available in Europe and elsewhere. It is the right combination of competitive and institutional funding that makes a system work. Where exactly that balance is, remains a hot topic of current debate.

Scientifically we have entered a new epoch of fundamental physics research. Thanks to gravitational wave experiments, EHT, ESO's near-infrared interferometer Gravity, but also pulsar observations we are now entering an era of experimental spacetime physics. Finally, we can test physics predictions that seemed impossible when they were first calculated. We have not found a new theory of gravity yet, but we have learned more about the physics in a world, where spacetime is torn and twisted unlike anything we can ever create here on earth. This has also led to an avalanche of new tests of GR and new ideas that are being explored. We are now standing on solid ground when we use general relativity in the most extreme circumstances. Maybe it is time to dare taking one step further and try to come up with a new theory that goes further than just Einstein. At least, we are now able to rule some of them out.

In the end, experiments not only measure, they also inspire. Hence, I hope that experiments like the Event Horizon Telescope will not only test old ideas, but also create room to grow new ones. Maybe the next Einstein is a little kid in Africa that was inspired to go into science because she saw an actual image of a black hole and wanted to know what lies beyond. Even in science dreams can sometimes become true.